\documentclass{svproc}
\usepackage{url}
\usepackage{graphicx}
\usepackage{amsmath}
\usepackage{bbm}
\usepackage{amsfonts}
\usepackage{amssymb}
\usepackage{latexsym}
\usepackage{xcolor}

\def\sfrac#1#2{{\textstyle{#1\over #2}}}
\newcommand{\be}{\begin{equation}}
\newcommand{\ee}{\end{equation}}
\newcommand{\ba}{\begin{array}}
\newcommand{\ea}{\end{array}}
\newcommand{\bea}{\begin{eqnarray}}
\newcommand{\eea}{\end{eqnarray}}

\def\nus{{\nu_s}}

\begin{document}
\mainmatter              
\title{Phantom Fluid Cosmology\\ {\it or}\\ Ghosts for Gordon}
\titlerunning{Phantom Fluid Cosmology}  
%
\author{James M.\ Cline}
\authorrunning{James M.\ Cline} 
%
%
\institute{McGill University, Montr\'eal, Qc, Canada\\
\email{jcline@physics.mcgill.ca},
\and
Trottier Space Insitute,  Montr\'eal, Qc, Canada}

\maketitle              

\begin{abstract}
The quanta of phantom dark energy models are negative energy
particles, whose maximum magnitude of energy $|E|$ must be less
than a cutoff $\Lambda\lesssim 20$\,MeV.  They are produced by
spontaneous decay of the vacuum into phantoms plus normal particles.
I review general cosmological constraints that have been derived 
from the effects of such phantom fluid production, and a possible 
application: the generation of boosted dark matter or radiation that
could be directly detected.  Recent excess events from the DAMIC 
experiment can be well-fit by such processes.

\end{abstract}
\section{Happy Birthday, Gordon!}
\begin{figure}[t]
\centerline{
\includegraphics[width=\hsize,angle=0]
{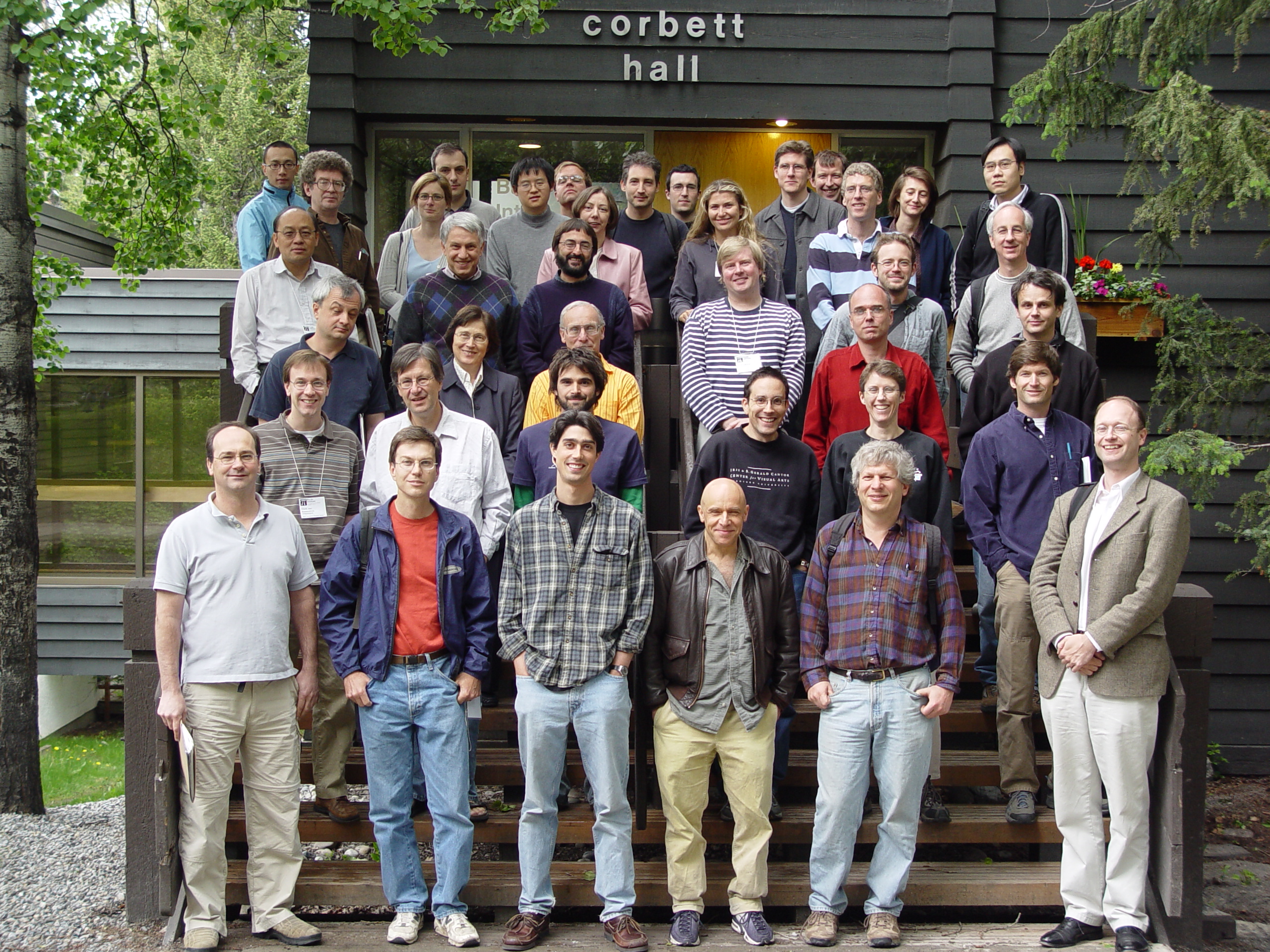}}
\caption{Participants of the 2004 BIRS workshop 
``New horizons in string cosmology.''
\label{fig:birs}}
\end{figure}

I never had the pleasure of collaborating with Gordon Semenoff, but
in 2004 we co-organized a conference at the Banff International Research
Station, called ``New horizons in string cosmology.''  The conference
photo, Fig.\ \ref{fig:birs}, has a peculiar feature, which I asked
the audience of the present meeting whether they could identify.
Namely, the four participants of this meeting who also attended
the BIRS meeting are all grouped together in the bottom left-hand
corner.  This seems to be a violation of thermodynamics, or
causality.\footnote{There was another peculiar feature which the
audience equally failed to notice: I was wearing exactly the same
clothes on both occasions, down to the glasses.  It was a little
difficult to see anything while giving the talk.}

For the present contribution, I tried to think of something that could
be remotely of interest to Gordon, considering our rather different
research paths, subsequent to our near overlap in the area of string
cosmology.  What follows is the best I was able to come up with.
It is a model of a new kind of dark energy, derived from a rather old
one, phantom scalar fields.  Unlike the old version, which relies upon
the condensate of a scalar field with a wrong-sign kinetic term, our 
version assumes the condensate is absent, but the corresponding
negative-energy particles, that are inevitably produced by the decay of the vacuum,
are present \cite{Cline:2023cwm}.  I will start with a review of the original phantom dark
energy models by way of motivation.

\section{Phantom dark energy: a brief history}
I was at CERN in 2003 on my first sabbatical when WMAP (Wilkinson
Microwave Anisotropy Probe) made its first data release.  Amongst
other things, these data led to constraints on the equation of state
of the dark energy that drives the acceleration of the Universe
\cite{WMAP:2003elm}, $w = p/\rho$,
where $p$ is the pressure and $\rho$ the energy density, with allowed
regions extending to values of $w < -1$, in violation of the dominant 
energy condition (DEC).  Theorists who had been
analyzing various combinations of cosmological data, including
pre-WMAP data of the cosmic microwave background, had also found 
such evidence, for example Refs.\ 
\cite{Hannestad:2002ur,Melchiorri:2002ux}.  

Even before these experimental hints, R.\ Caldwell had pointed out a
simple way of violating the DEC in the context of scalar field
theories \cite{Caldwell:1999ew}.  Recall that a cosmological constant
with stress-energy tensor $T^{\mu\nu} = \Lambda\,{\rm diag(1,-1,-1,-1)}$
saturates the DEC, while a homogeneous scalar field has equation 
of state
\be
	w = {\sfrac12\dot\phi^2 -V\over 
	\sfrac12\dot\phi^2 + V} \ge -1\,,
	\label{weq}
\ee
where the inequality holds assuming that $V>0$.  This interpolates
between cosmological constant, in the limit $\dot\phi\to 0$, and 
$w=1$ when $V\to 0$.  Caldwell's idea was to consider a scalar field
with a wrong-sign kinetic term, a ``phantom'' field, such that
$\dot\phi^2\to -\dot\phi^2$ in Eq.\ (\ref{weq}) (but $V$ remains $\ge
0$).  Then $w < -1$ if $\dot\phi^2 < 2V$, and diverges to $-\infty$
as $\dot\phi^2 \to 2V$.   This will lead to a ``big rip''
singularity within a finite time, in which the expansion rate of the
universe diverges and gravitationally bound structures are ripped
apart.\footnote{It is interesting that Caldwell's paper, although
written in 1999, was not published until 2002, when CMB data were
hinting at the possibility of $w<-1$.}  This is readily understood
from the classical equation of motion of a homogeneous phantom field,
\be
	\ddot\phi + 2 H \dot\phi = +V'(\phi)\,;
\ee
the field climbs {\it up} the potential instead of rolling down.

Currently there is no compelling evidence for violation of the DEC;
the Particle Data Group gives $w = -1.028 \pm 0.031$ 
\cite{Workman:2022ynf}.  However, this has not stopped theorists from
continuing to consider phantom dark energy.  InSpire database gives 
1931 results, showing a steady rate of publication over the last two
decades.  The vast majority of these papers consider the phantom as a
classical field.  

But phantoms are not exempt from quantum mechanics,
and one should ask what is the quantum field theoretic consequence of
a wrong sign kinetic term.  The answer depends on the choice of sign
for the $i\epsilon$ prescription in the propagator,
\be
	P(p) = {-i\over p^2 - m^2 \pm i\epsilon}\,.
\ee
The positive sign leads to negative probability as in the case of
Faddeev-Popov ghosts, whereas the negative sign corresponds to
positive probability, but with negative energy particles propagating
into the future.  The latter choice is the lesser of the two evils,
and so one must consider phantom particles to carry negative energy if
they are to be taken seriously as quantum fields.

\begin{figure}[t]
\centerline{
\raisebox{2cm}{\includegraphics[width=0.3\hsize,angle=0]{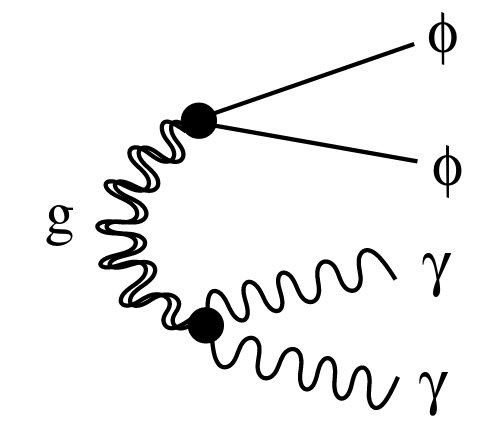}}
\includegraphics[width=0.7\hsize,angle=0]{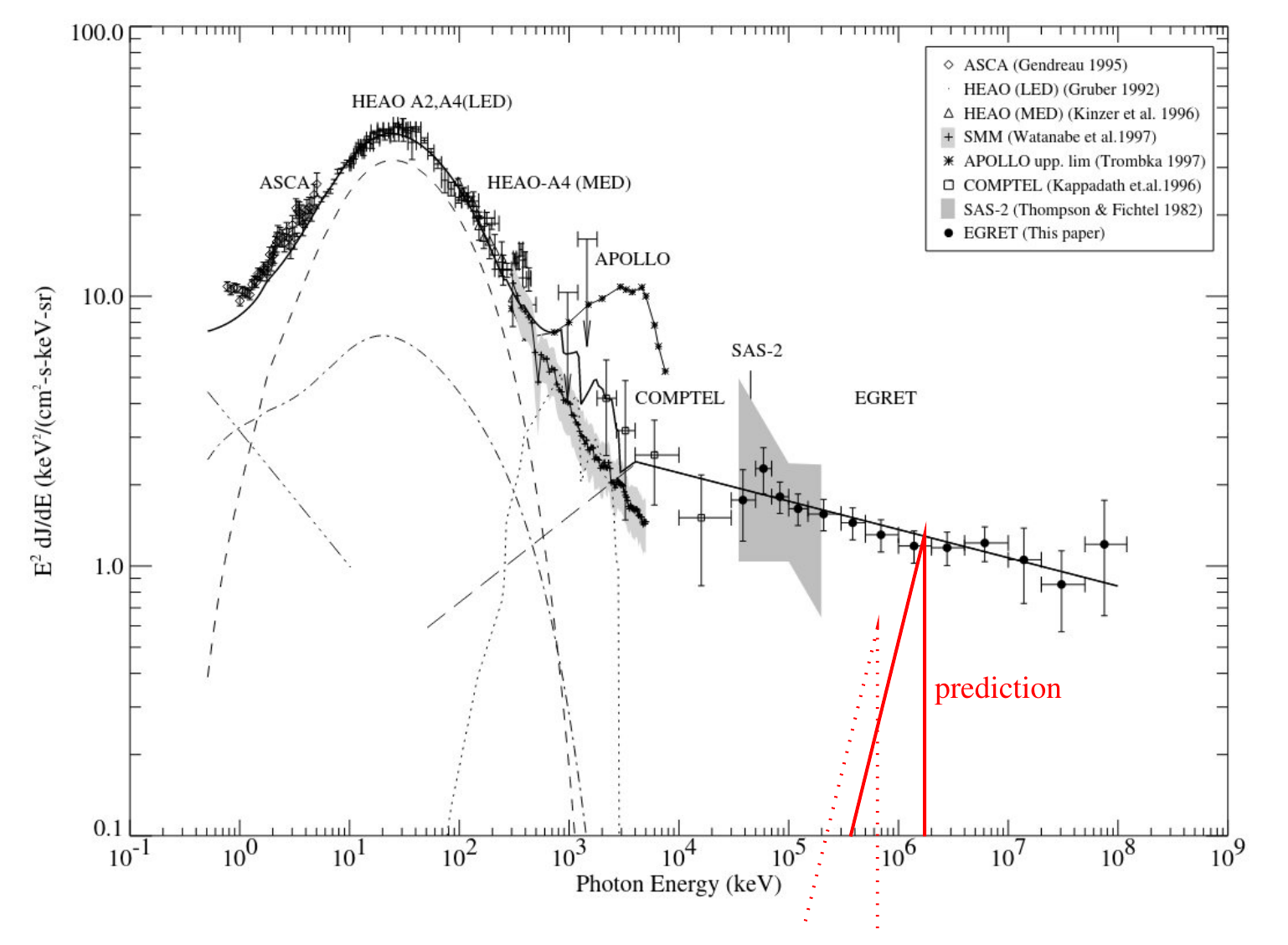}}
\caption{Left: Feynman diagram for decay of the vacuum to two ghosts
plus two photons.  Right: schematic illustration of predicted x-ray
spectrum (for two values of $\Lambda$, in red) versus cosmological diffuse photon background.
\label{fig:vac-decay}}
\end{figure}

Ref.\ \cite{Carroll:2003st} understood that the phase space of phantom
particles must be cut off at some scale $\Lambda$ to avoid divergent decay rates, since
arbitrarily negative energy phantoms could otherwise appear in the final
state.  But it was Ref.\ \cite{Cline:2003gs} that pointed out the most stringent
irreducible constraint, arising from decay of the vacuum into two
phantoms plus two photons, mediated by gravity.  The Feynman diagram
is shown in Fig.\ \ref{fig:vac-decay}.  On dimensional grounds, it
leads to a constant decay rate per unit volume of order $\Gamma \sim 
\Lambda^8/ m_P^4$, where $m_p$ is the (reduced) Planck mass.
The ensuing photon spectrum roughly takes the form
\be
{dn\over dE} \sim {\Lambda^7 t_0\over m_p^4}\,\Theta(\Lambda-E)\,,
\ee
where $t_0$ is the age of the Universe.  It rises steeply in
amplitude with $\Lambda$, but only linearly in energy.  The maximum
value of $\Lambda$ such that the prediction lies below the observed 
diffuse photon background was estimated to be $\sim 3\,$MeV, which is
rather low for an ultraviolet cutoff.  Above this scale, the phantom
description must break down and revert to a more normal-looking field
theory.

\section{Revisiting phantom fluids} 
Since 2003, there have been a few
developments motivating a second look at the cosmological production
of ghost particles.  One is the  Hubble tension: the discrepancy
between the CMB determination of the Hubble constant, and that coming
from more local measurments such as type Ia supernovae; see Ref.\
\cite{DiValentino:2021izs} for a review.  My naive idea was to suppose that the vacuum
could decay into massless ghosts plus dark matter.  Due to the faster
redshifting of the ghost energy density, there would be a net gain in
dark matter as a function of time, which could explain why the
Universe expands more rapidly at late times than expected.  But this
should not be a viable resolution, as it has been widely understood
that late dark energy solutions do not work for the Hubble tension,
while early dark energy may be able to do so.  The current mechanism
is late-time since the bulk of the accumulated energy is appearing
today.  The problem with the naive argument is that the appearance of
the CMB depends upon the entire expansion history of the Universe, and
not just that at early times.

Dark matter produced by vacuum decay to ghosts must be light,
with a mass below the $\sim$MeV cutoff.  At the same time, it will be
produced with an energy of order $\Lambda$, hence it is boosted to
much higher speeds than ordinary DM, which virializes in the galaxy
with speed $\sim 10^{-3}c$.  This is reminiscent of another recent
development, the interest in boosted light DM \cite{Agashe:2014yua}.
Such light DM particles could not be directly detected if moving
slowly, being below the energy threshold of conventional detectors, but boosted
DM is able to deposit sufficient energy.  Interest in
such models was further stimulated by excess events (now attributed to
background) observed by the XENON1T experiment in 2010
\cite{XENON:2020rca}.  
These were the motivations for our recent work on the subject, 
Refs.\ \cite{Cline:2023cwm,Cline:2023hfw}.

\section{Revised limit on $\Lambda$}
\begin{figure}[t]
\centerline{
{\includegraphics[width=0.5\hsize,angle=0]{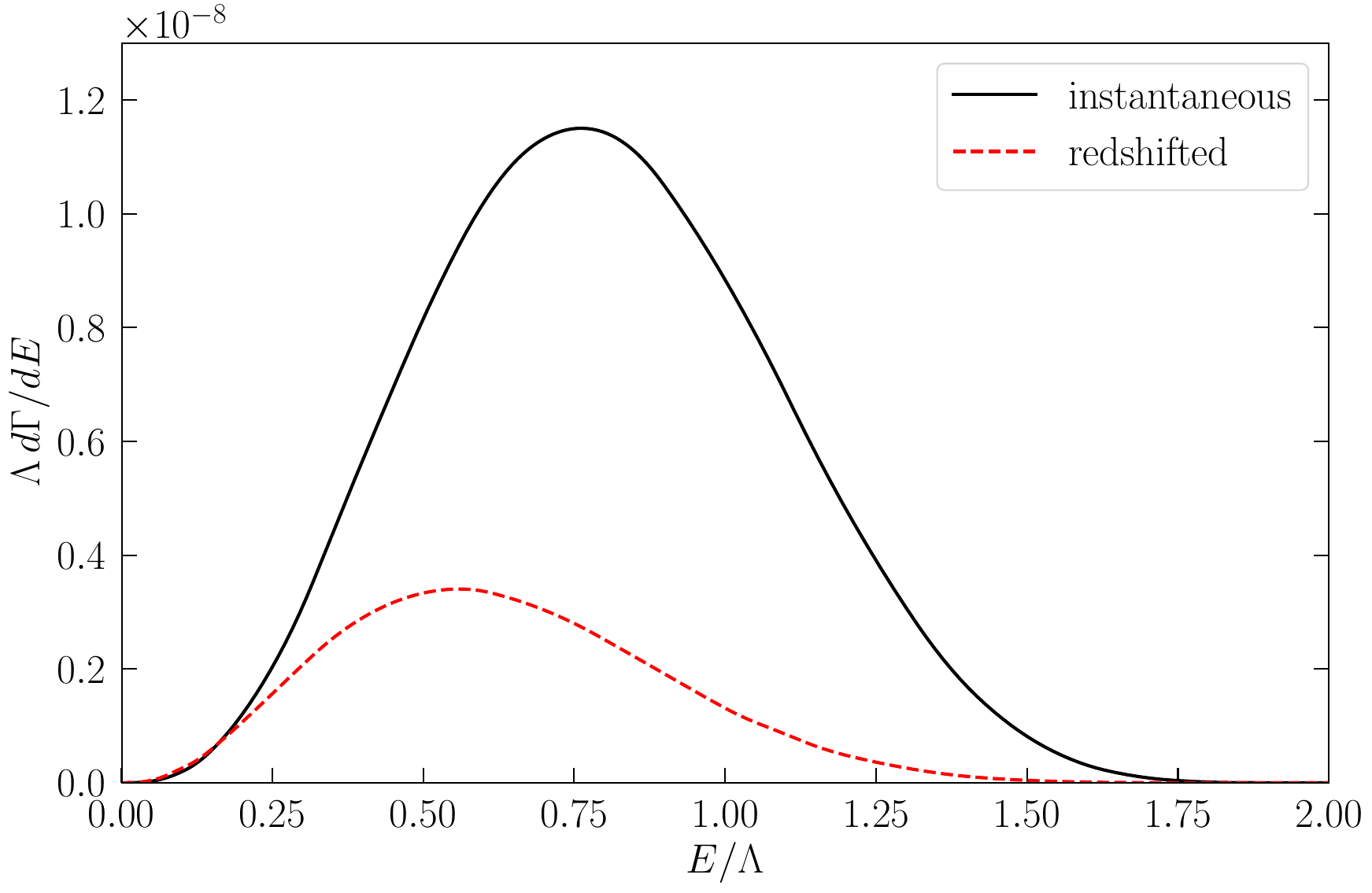}}
\includegraphics[width=0.5\hsize,angle=0]{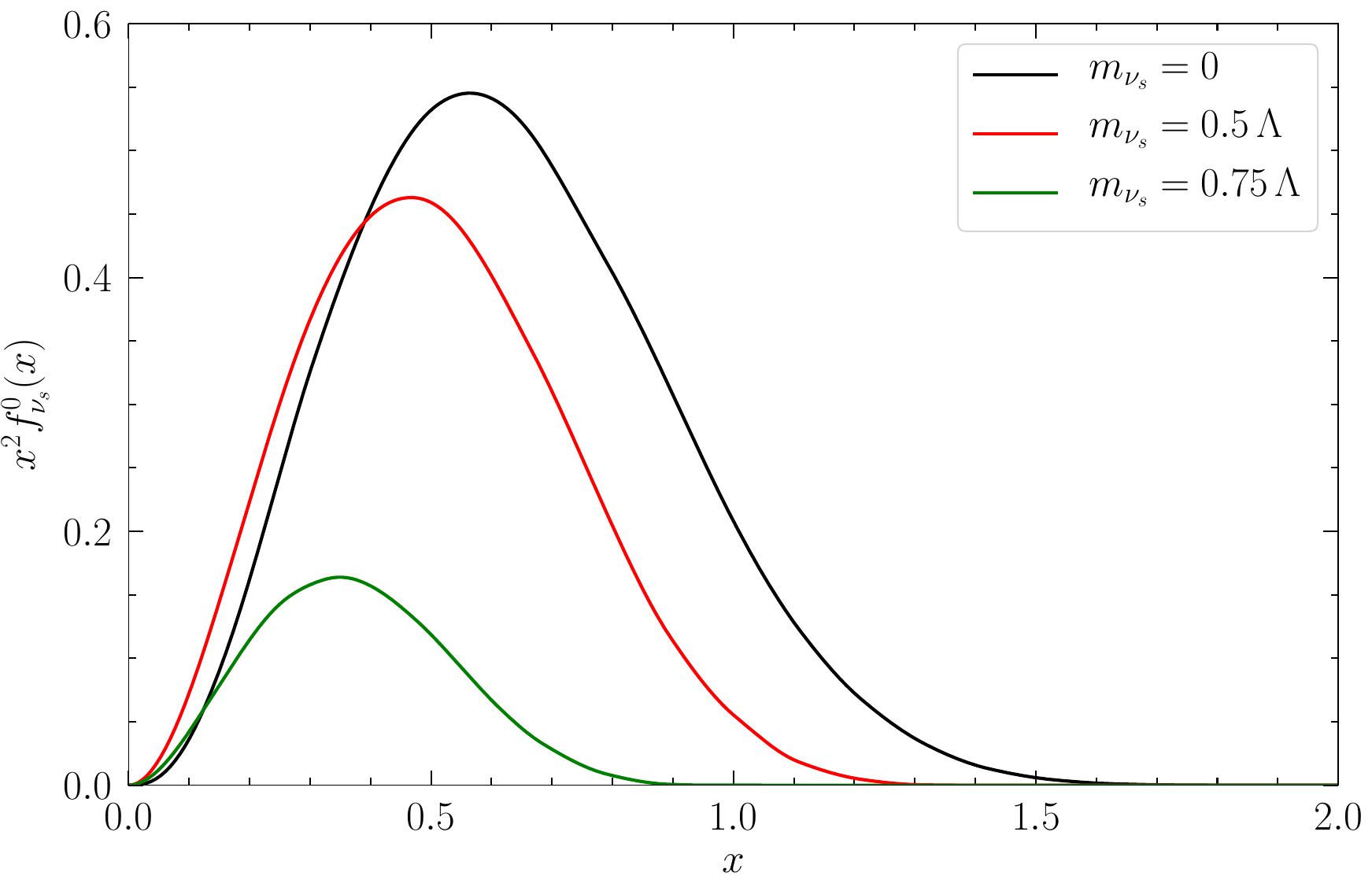}}
\caption{Left: spectrum of photons from decay of the vacuum, to
$\gamma\gamma\phi\phi$. Right: spectra of sterile neutrinos from
vacuum decay to $\nu_s\bar\nu_s\phi\phi$.
\label{fig:spectra}}
\end{figure}

One of our goals was to better quantify the limit on $\Lambda$ from vacuum
decay into ghosts plus photons, improving on the original
order-of-magnitude estimate.  First, we used the exact expression for
the matrix element, which takes the rather aesthetic form
$|{\cal M}|^2 = (t u/s)^2/8 m_p^4$ \cite{Bernal:2018qlk}.  Second,
we computed the phase space integral accurately to obtain the full
photon spectrum.  Third, we accounted for reshifting of the spectrum
using an appropriate Boltzmann equation, with the result shown in Fig.\
\ref{fig:spectra}.  Finally, we used more recent
observational data from the COMPTEL experiment.  The new limit on the
cutoff is
\be
	\Lambda < 18\,{\rm MeV}\,,
\label{Lbound}
\ee
six times less stringent than the original estimate. 

\section{Contstraints on phantom fluid production}

For definiteness, we consider a sterile neutrino $\nu_s$ as the light
DM candidate coupled to phantoms, but similar results would hold for
scalar DM.  However the gravitational coupling, analogous to Fig.\
\ref{fig:vac-decay}, turns out to be too weak (by some 9 orders of
magnitude) to be cosmologically relevant.  Therefore we consider the
possibility that the decay to $\phi\phi\nus\bar\nus$ is mediated by
a more strongly coupled intermediate particle, such as a heavy vector
or scalar boson.  For definiteness I will assume the vector case here
(scalar exchange gives similar results); 
by integrating out the heavy particle, it can be
described by an effective interaction,
\be
	{\cal L} \quad\ni\quad 
{i\over M_v^2}(\phi^*\!\!\stackrel{\leftrightarrow}{\partial}_\mu\!\phi)\,
	\bar\nus\gamma^\mu\nus\,,
\label{EFT}
\ee
where the ghost is complex and the neutrino is Dirac, consistent with
a U(1) gauge interaction.  We will show that $M_v$ as large as the GUT
scale can lead to observable effects, if $\Lambda$ saturates the bound
(\ref{Lbound}).

\begin{figure}[t]
\centerline{
{\includegraphics[width=0.5\hsize,angle=0]{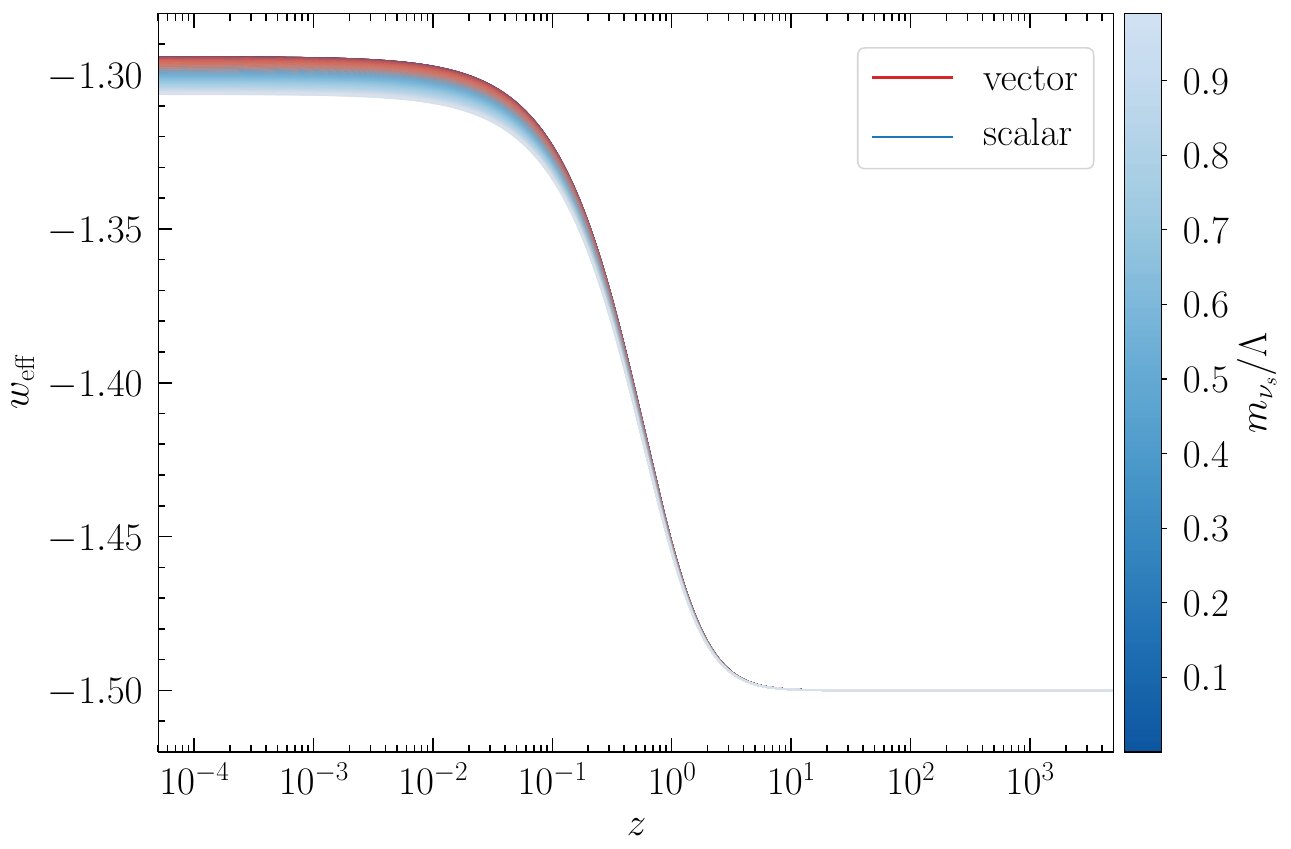}}
\includegraphics[width=0.5\hsize,angle=0]{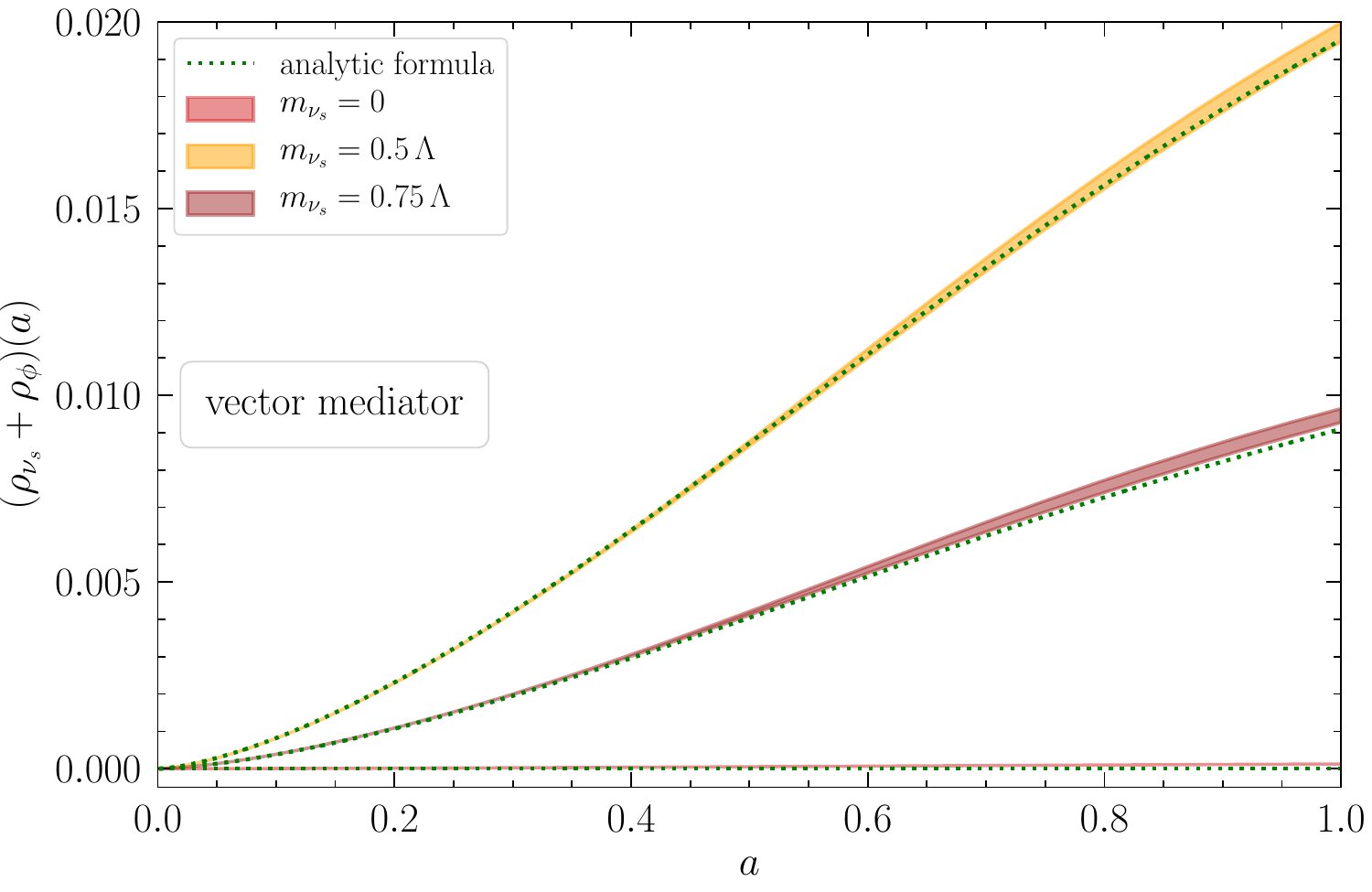}}
\caption{Left: equation of state of the ghost fluid as a function of
redshift.  Band indicates variation with respect to $m_\nus/\Lambda$
for vector and scalar mediator models.  Right: growth of the phantom
fluid energy density with scale factor, for several $m_\nus/\Lambda$
values.
\label{fig:eos}}
\end{figure}

For simplicity we take the phantoms to be massless.  Then there are
two qualitatively different regimes: $\nus$ is massive or (nearly)
massless.  The first case is straightforward since it doesn't require
cosmological perturbation theory.  Instead the main effects are via
the time-dependent energy densities of the ghosts and $\nus$s produced
from vacuum decay.  They are described by coupled Boltzmann equations,
\bea
	{d\rho_{\phi}\over dt} + 4H \rho_{\phi} &=& -\Gamma_\rho, \quad
	{d\rho_{\nus}\over dt} + 3H (1 + w_{\nus}) \rho_{\nus} = +\Gamma_\rho\,,
\label{boltzeqs}
\eea
where the rate of energy density production of $\nu_s$ is
\be
		\Gamma_\rho = \int\prod_i {d^{\,3}p_i\over (2\pi)^3 2|E_i|} 
	(2\pi)^4\delta^{(4)}\Big(\sum_i p_i^\mu\Big)|{\cal M}|^2 \cdot (E_{\nus} + E_{\bar\nus})
\ee
with matrix element $|{\cal M}|^2 = (2/M_v^2)(s^2 - (t-u)^2)$, 
and the Hubble parameter is
\be
	H^2 = H_0^2\times(\rho_{\rm rad} + \rho_m + \rho_\Lambda + 
	{\color{red} \rho_\phi +
	\rho_{\nus}})/\rho_c\,,
\ee
where $\rho_c$ is the critical density.
Because of the terms in red, the Boltzmann equations (\ref{boltzeqs})
are nonlinear, but in practice, their sum must remain relatively small
to avoid undue modifications of standard cosmology, and then simple
analytic approximate solutions can be found.

One can see that if $\nus$ is massless, $w_{\nus} = 1/3$, then the
phantom fluid energy density 
$\rho_g\equiv \rho_\phi + \rho_{\nus}=0$ is an exact, rather uninteresting solution.
But if $m_\nus>0$, the two species redshift differently, leading to
net production of $\rho_g$.  A potential complication is that
$w_{\nus}$ could be time-dependent, since the $\nus$s are being produced
at a constant rate:
\be
	w_{\nus} = {p_{\nus}\over\rho_{\nus}} = {\int d^{\,3}p\,f_{\nus}(p,t)\, p^2/3E\over
	\int d^{\,3}p\,f_{\nus}(p,t)\, E} \sim \frac13 
-\left(m_{\nus}\over\Lambda\right)^2 +\cdots
\ee
hence their distribution function $f_\nus$ is time-dependent.  However
we find that this dependence cancels nearly exactly in the ratio, so
that $w_\nus$ is only a function of $m_\nus/\Lambda$, which greatly
simplifies the analysis.

In effect, $\rho_g$ behaves as an exotic form of dark energy,
different from that of the classical phantom condensate, with
effective 
equation of state $w_{\rm eff} = (p_\phi+p_\nus)/(\rho_\phi+\rho_\nu)$
as shown in Fig.\ \ref{fig:eos} (left).   It violates the DEC,
$-1.5\le w_{\rm eff} \lesssim -1.3$.  Hence, unlike
matter, $\rho_g$ grows with time, as shown in Fig.\ \ref{fig:eos}
(right).

\begin{figure}[t]
\centerline{
{\includegraphics[width=0.5\hsize,angle=0]{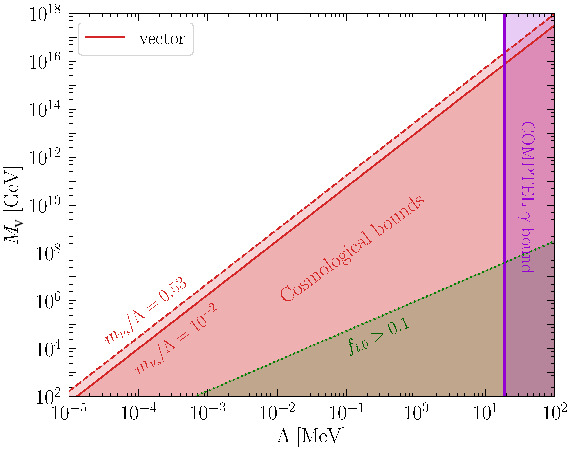}}
\includegraphics[width=0.5\hsize,angle=0]{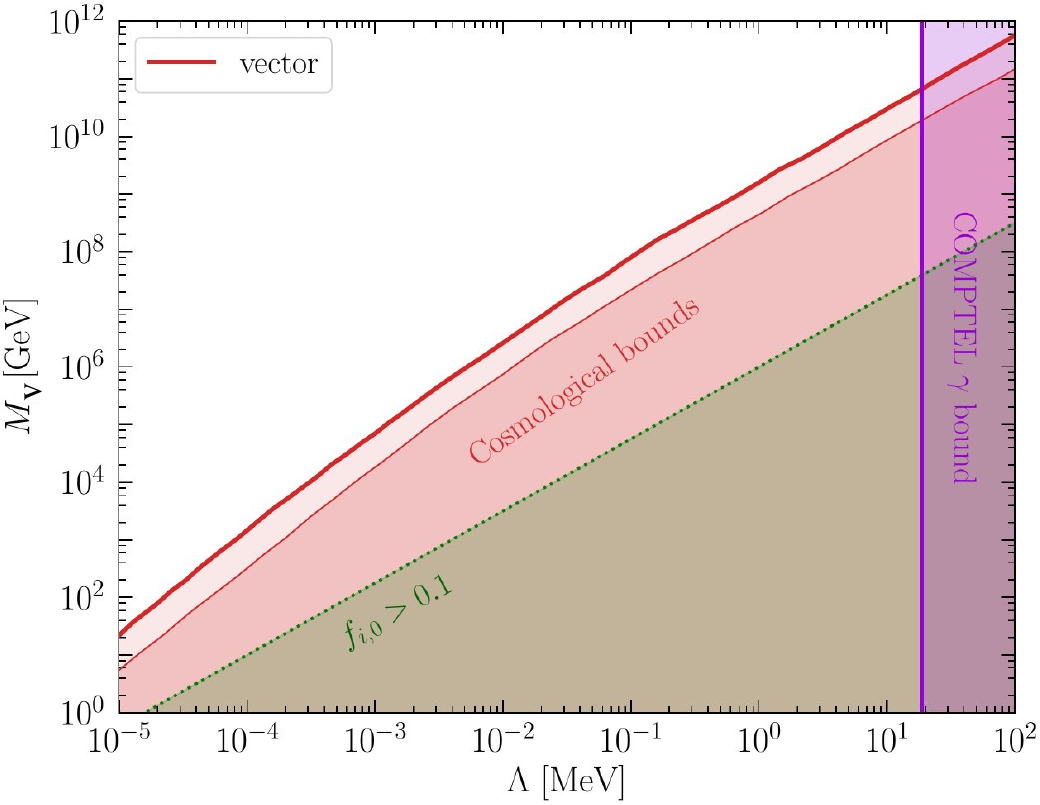}}
\caption{Left: Lower limit on the new physics scale $M_v$ versus
$\Lambda$ from phantom fluid production, for $m_\nus>0$.  Right: same
for $m_\nus=0$.  Brown regions are excluded by the technical
assumption that the fluids do not become degenerate (for which our
analysis is not valid).
\label{fig:const}}
\end{figure}

For the massless $\nus$ case, no constraint arises at zeroth order in cosmological
perturbation theory, even with arbitrarily
fast growth of $\rho_\phi$ or $\rho_\nus$ individually, since they
cancel exactly.  However, their perturbations do not exactly cancel,
since their distribution functions are different (Bose-Einstein versus
Fermi-Dirac) and their phase space limits differ (only ghosts are
restricted to have $|\vec p|<\Lambda$).  Hence their first-order
perturbations can grow and impact the integrated Sachs-Wolfe effect and lead to
constraints from the cosmic microwave background (CMB).  

\begin{figure}[t]
\centerline{
\includegraphics[width=0.5\hsize,angle=0]{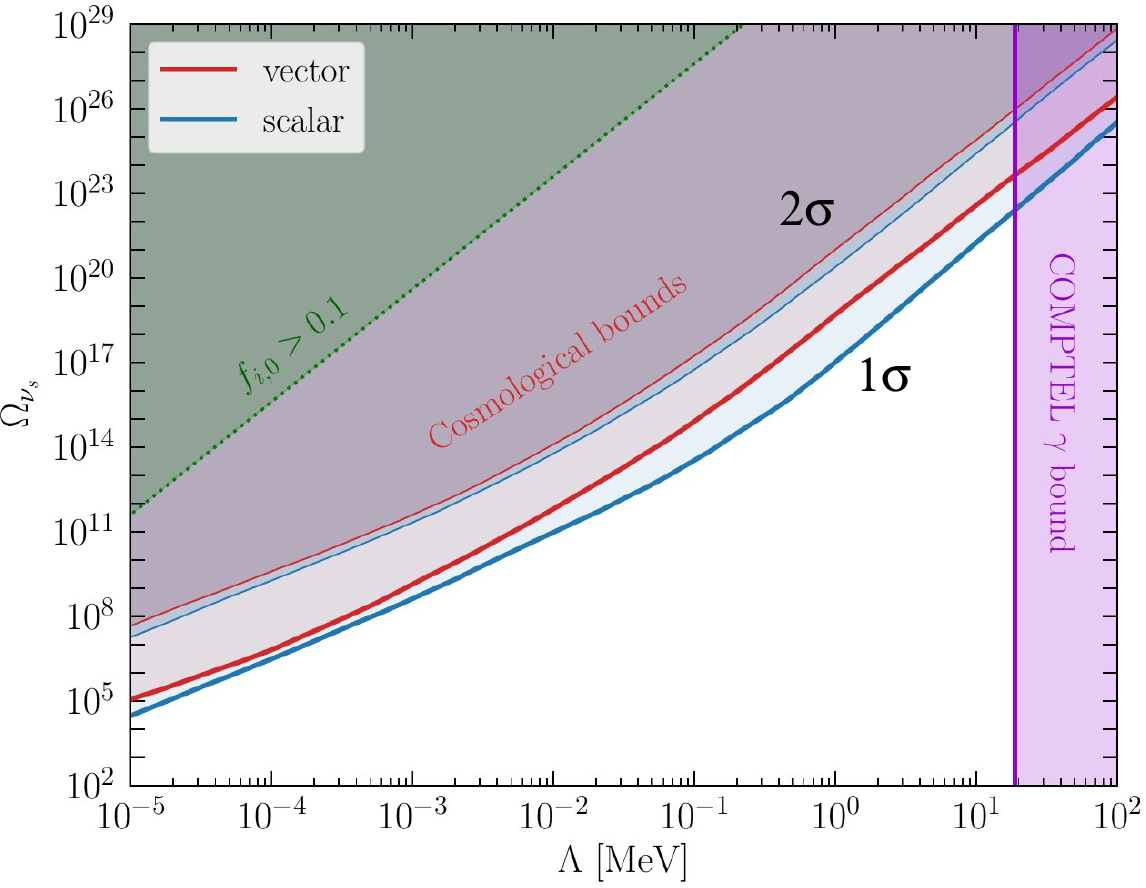}
\includegraphics[width=0.56\hsize,angle=0]{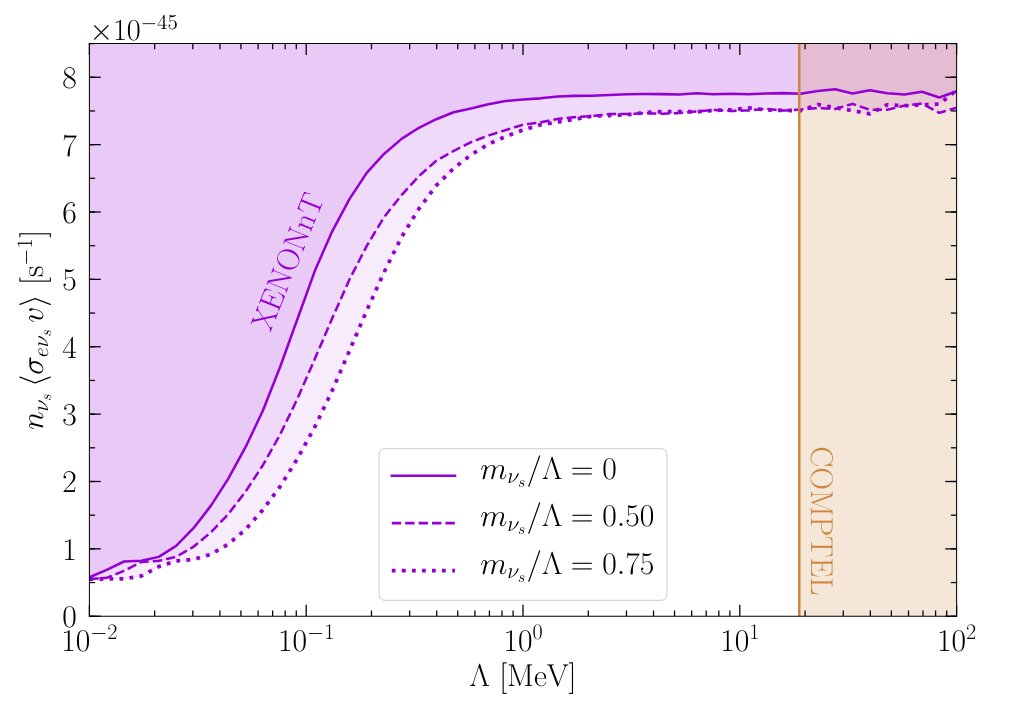}
}
\caption{Left: upper limit on $\Omega_\nus$ versus $\Lambda$ for the case of
massless $\nus$.  Right: direct detection upper limit on cross section
for $\nus$-$e$ scattering times $\nus$ density versus $\Lambda$, from 
XENONnT null results.
\label{fig:meq0}}
\end{figure}

For both massive and massless $\nu_s$, we
performed a Monte Carlo search of the parameter space using CosmoMC
\cite{Lewis:2002ah} to derive limits from combined cosmological data
sets on the new physics parameter $M_v$, as well as on 
the phenomenological quantity $\Omega_g$ (the present contribution
of the ghost fluid to the cosmological density) as functions of the 
cutoff scale $\Lambda$.  $\Omega_g$ cannot exceed $0.2$, so
disappointingly it is not a viable replacement for the cosmological
constant $\Omega_\Lambda \gtrsim 0.5$, and could only provide a
fraction of the total dark energy.  The ensuing lower limits on
$M_v$ versus
$\Lambda$ are shown in Fig.\ \ref{fig:const} for massive $m_\nus$.
For $m_\nus=0$, the density of $\nu_s$ can grow to enormous values,
$\Omega_\nus\sim 10^{25}$ since it is canceled by the ghosts.  The
limit is shown in Fig.\ \ref{fig:meq0} (left).

We investigated the extent to which the phantom fluid model can
ameliorate the Hubble tension, as well as the $S_8$ tension ($S_8$
being the parameter describing the amplitude of fluctuations at the
scale $8 h^{-1}$ Mpc).  As expected on general grounds, this being a
late-dark energy model, it does not provide a full resolution, but it
reduces the discrepancy to the $3\sigma$ level.  The main source of
the tension is the baryon acoustic oscillation (BAO) data.  If one
omits these sets, excellent consistency can be obtained.

\section{Direct detection}
With the possibility of such a huge exotic dark radiation component,
one is led to wonder if it could have observable consequences. 
Suppose it was coupled to electrons in a similar manner to Eq.\
(\ref{EFT}), 
\be
	{\cal L} \quad\ni\quad 
{i\over M'^2}(\bar e \gamma_\mu e)
	(\bar\nus\gamma^\mu\nus)\,,
\ee
where we allow for a different interaction strength $M'$.  We choose
electrons since recent experimental searches for light dark matter
have been taking advantage of $e$-DM scattering: it is easier for a
light particle to impart energy to electrons than to nuclei.  The dark
radiation of massless $\nu_s$ carries energy $\sim\Lambda$ which can
be far greater than that of thermal radiation, and easily above the
energy thresholds $\sim 20$\,eV of detectors such as DAMIC 
\cite{DAMIC:2023ela}.
In Ref.\ \cite{Cline:2023hfw} we calculated the event rate for such interactions in
xenon and silicon detectors.  A model-independent constraint on the
scattering cross section of $\nus$ on electrons times the $\nus$
density can be derived from the null results of the XENONnT
experiment \cite{XENON:2022ltv}, which exclude the putative signal that was previously 
observed by XENON1T \cite{XENON:2020rca}.  It is shown in Fig.\ \ref{fig:meq0} (right). 

\begin{figure}[t]
\centerline{
\includegraphics[width=0.5\hsize,angle=0]{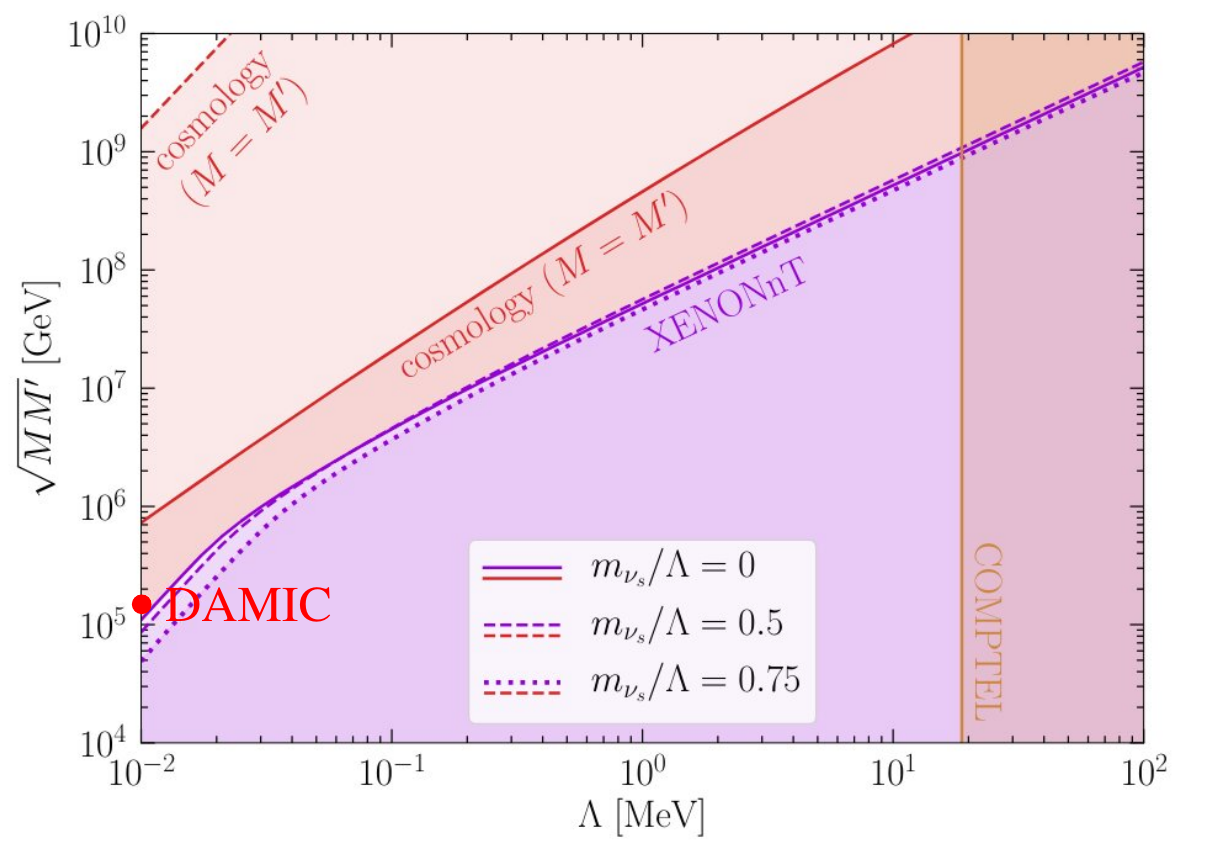}
\includegraphics[width=0.52\hsize,angle=0]{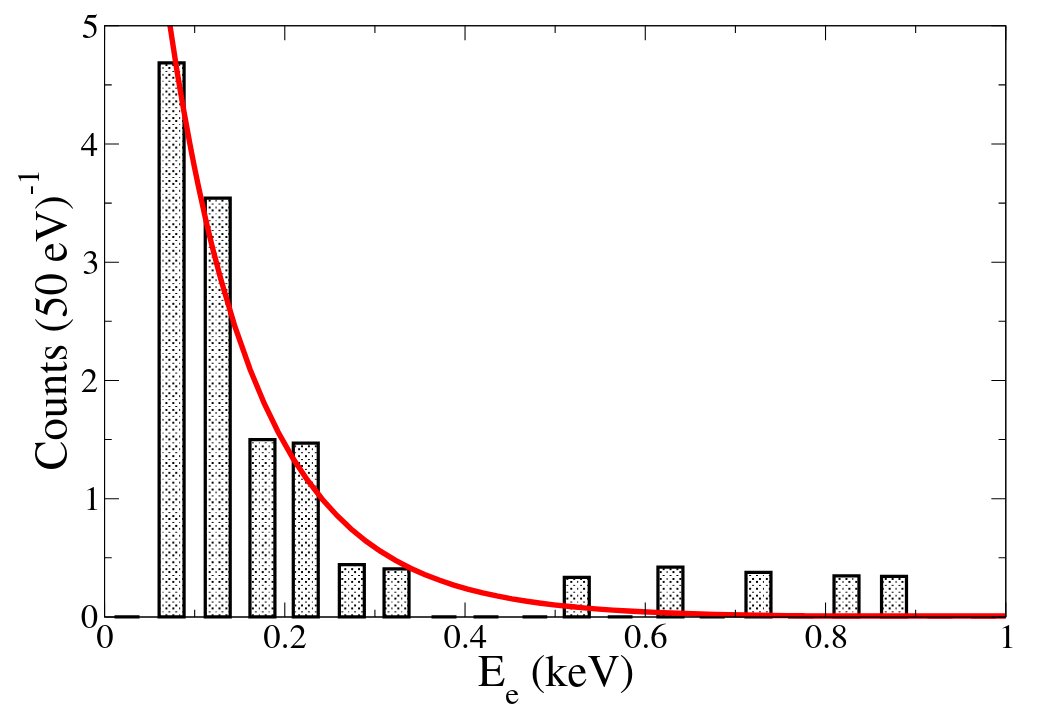}
}
\caption{Left: XENONnT limit on $\sqrt{M_v M'}$, compared to previous
cosmological limits assuming $M'=M_v$.  Right: massless $\nus$ fit to the DAMIC
excess signal, with $\Lambda = 10\,$keV and $\sqrt{M'M_v} = 1.3\times
10^5$\,GeV. 
\label{fig:final}}
\end{figure}

Translating the XENONnT constraint into model parameters, we find 
an upper bound on the product $M_v M'$, corresponding to the product
$\sigma_{\nus e}n_\nus$.  The signal turns out to be undetectably 
small for cosmologically allowed values if we assume $M'=M_v$; one
needs $M' > M_v$ to saturate the bound.  This is illustrated in Fig.\
\ref{fig:final} (left).  The DAMIC experiment in SNOLAB has observed
excess events that can be interpreted as scattering on electrons, with
a spectrum shown in Fig.\  \ref{fig:final} (right).  

The phantom fluid
model has essentially two parameters relevant for direct detection: $\Lambda$, determining
the shape of the spectrum and
$\sqrt{M_vM'}$, determining its normalization; 
the value of $m_\nus/\Lambda$ has a relatively small effect.  We find
an excellent fit to the observed data by taking $\Lambda = 10\,$keV
and $\sqrt{M_v M'} = 1.3\times 10^5\,$GeV, which is not far from the 
XENONnT lower limit.  It must be raised to the 8th power to get
$\sigma_{\nus e}n_\nus$, implying the signal is $1.3^8\sim 8$ times 
lower than the current XENONnT sensitivity.  There is a data bin at 
0.05\,keV that is essentially empty.  We confirm that the predicted 
signal falls to negligible values in this bin, due to the binding
energy of silicon preventing excitations.

\section{Conclusions}  

Phantom fluids through vacuum decay are an inevitable
consequence of phantom dark energy models, although it is necessary to couple the
ghosts more strongly than gravity to make them cosmologically relevant.  If one
does this, for (massless) ghosts and light dark matter of mass $\lesssim 18$\,MeV, it gives
an exotic dark energy component with equation of state $-1.5< w< -1.3$ that
violates the dominant energy condition.  Sadly, cosmological data prevents it
from contributing more than $\Omega_g\sim 0.2$ to the total energy budget. 
Nevertheless, even a small amount will lead to a regularized big rip in the
future---regularized, since the theory must revert to being normal at scales above
the cutoff.  $\Lambda \sim 10$\,keV is sufficient to rip apart bound
objects like the sun in a future epoch.  Apart from these annoying features, phantom fluids can also do something useful
by providing a new target for detection of boosted light dark matter or radiation.

This work was supported by NSERC (Natural Sciences and Engineering Research
Council, Canada).

\end{document}